\documentclass[aps,prb,superscriptaddress,floatfix]{revtex4}
\usepackage{epsfig,amsmath,amssymb,color}
\def\beq{\begin{equation}}
\def\eeq{\end{equation}}
\def\bea{\begin{eqnarray}}
\def\eea{\end{eqnarray}}
\def\nn{\nonumber}

\begin{document}

%\tolerance 50000 \preprint{
%\begin{minipage}[t]{1.8in}
%\rightline{La Plata-Th 00/xx} \rightline{}
%\end{minipage}
%}

%%%%%%%%%%%%%%%%%%%%%%%%%%%%%%%%%%%%%%%

\title{Numerical Jordan-Wigner approach for two dimensional spin systems}

\author{D.C.\ Cabra$^{1,2}$ and G.L.\ Rossini$^2$}

\affiliation{
~~\\
$^1$Laboratoire de Physique Th\'eorique\\
Universit\'e Louis Pasteur\\
3 rue de l'Universit\'e, F-67084 Strasbourg Cedex, France\\
~~\\
$^2$Departamento de F\'{\i}sica\\
Universidad Nacional de la Plata\\
C.C.\ 67, (1900) La Plata, Argentina\\
and \\
Facultad de Ingenier\'\i a, \\
Universidad Nacional de Lomas de Zamora,\\
Cno. de Cintura y Juan XXIII, (1832) Lomas de Zamora, Argentina.}
\date{\today}

\begin{abstract}

\begin{center}
\parbox{14cm}{
We present a numerical self consistent variational approach based on the Jordan-Wigner
transformation for two dimensional spin systems. We apply it to the study of the well
known quantum ($S=1/2$) antiferromagnetic $XXZ$ system as a function of the easy-axis
anisotropy $\Delta$ on a periodic square lattice. For the $SU(2)$ case the method
converges to a N\'eel ordered ground state irrespectively of the input density profile
used and in accordance with other studies. This shows the potential utility of the
proposed method to investigate more complicated situations like frustrated or disordered
systems.}
\end{center}
\end{abstract}

\maketitle

\section{Introduction}

Quantum spin systems in two dimensional (2D) lattices have been the subject of intense
research, mainly motivated by their possible relevance in the study of high temperature
superconductors \cite{Anderson}. On the other hand, high magnetic field experiments on
materials with a 2D structure which can be described by the Heisenberg antiferromagnetic
model in frustrated lattices have revealed novel phases as plateaux and jumps in the
magnetization curves \cite{SrCuBo}. In spite of the huge efforts made, a general
understanding of the phase diagram of such magnets is elusive and it is then worth trying
to develop new techniques to study these systems systematically. Among the many different
techniques that have been used to study such systems, the generalization of the
celebrated Jordan-Wigner (JW) transformation \cite{JW} to two spatial dimensions
\cite{Fradkin} has some appealing features. It allows one to write the spin Hamiltonian
completely in terms of spinless fermions in such a way that the $S=1/2$ single particle
constraint is automatically satisfied due to the Pauli principle, while the magnetic
field enters as the chemical potential for the JW fermions. The price one has to pay is
the appearance of complicated non-local interactions between fermions. This method has
been applied in \cite{ARF} (see also \cite{Wang}) to study the $XXZ$ Heisenberg
antiferromagnet. These studies have been reviewed in \cite{Derzhko}.

More recently this technique was used to obtain a theoretical magnetization curve for the
Shashtry-Sutherland model, reproducing at the mean field level some of the experimentally
observed features for the material SrCu$_2$(BO$_3$)$_2$ which is assumed to be described
by such model \cite{Misguich}. Also the $J_1-J_2$ model, in relation to Li$_2$VOSiO$_4$
and Li$_2$VOGeO$_4$ compounds \cite{Chang}, and the $XY$ model \cite{Derzhko2003} were
analyzed with the same technique. All the studies performed have been based on a mean
field decoupling scheme as the starting point to deal with the non-local interactions
introduced by the JW transformation. In \cite{ARF} the mean field procedure was further
supplemented by the inclusion of fluctuations in terms of an auxiliary gauge field with a
leading Chern-Simons dynamics coupled to the lattice fermions. However, in spite of the
partial success of the JW transformation, many problems remain open, in particular in
connection to the study of frustrated systems such as the triangular lattice, etc. In
some cases, the results obtained via a direct mean field treatment lead to results that
are believed to be incorrect, like the appearance of a spin gap in the triangular lattice
case (see the discussion in \cite{Misguich}).  The main problem associated with the JW
approach is related to the implementation of the above mentioned mean field decoupling,
which renders the description approximate. Another highly non-trivial problem is the
construction of the lattice description of the Chern-Simons theory, which has been
carefully studied for the square lattice case only\cite{Semenoff}.

It is the purpose of the present paper to propose a systematic self consistent
mean field method for exploring the ground state (GS) of 2D lattice spin 1/2 systems, in a
way that could be applied to arbitrary lattice topologies. The method can
also be used in the presence of an external magnetic field, at finite
temperature and even be applied to disordered systems.

\section{Jordan-Wigner transformation in  $2$-dimensions}

The Jordan-Wigner transformation in $2$ spatial dimensions was originally
proposed in \cite{Fradkin} as a generalization of the well known
transformation in 1D, and has been further developed in \cite{ARF,Wang}.
It
maps a set of spin 1/2 operators $\vec{S}_p$ on lattice sites $p$ into
spinless fermion operators $c_p$ by
\bea
    S^-_{p} & = & c_p \exp \left[i \sum_{q\neq p} \theta_{qp} c^\dagger_q
c_q \right]\ ,
    \nn \\
    S^+_{p} & = & c^\dagger_p \exp \left[-i \sum_{q\neq p} \theta_{qp}
c^\dagger_q c_q \right]\ ,
    \nn\\
    S^z_p &=& c^\dagger_p c_p -1/2 \ ,
    \label{jw}
\eea
where $S^\pm = S^x \pm i S^y$ are the usual spin
raising and lowering operators and $\theta_{qp}$ is the argument of the
vector drawn from site $p$ to site $q$. The transformation is non-local,
and sets a preferred quantization axes $z$. The spin operators (\ref{jw})
satisfy bosonic $SU(2)$ commutation relations, while the Pauli principle
ensures that they belong to the irreducible representation $S = 1/2$.
Indeed, the only necessary ingredient that ensures the $SU(2)$ commutation
relations is the assignment of the phase factors which satisfies, for each
pair of sites $p,q$
\beq
    e^{i \theta_{pq}}\, e^{-i \theta_{qp}}=-1.
\label{key}
\eeq
One
should notice that there is a large freedom in choosing phase factors
satisfying this condition (\ref{key}). For instance, one could arbitrarily
shift $\theta_{pq} \to \theta_{pq}+2k\pi$ with different integers $k$ for
each pair of lattice points $p,q$, or even perform an arbitrary
simultaneous rotation for $\theta_{pq}$ and $\theta_{qp}$.  Standard plane
angles $-\pi< \theta \leq \pi$ measured from the $x$ axis is just the
simplest translation invariant choice on the flat infinite plane. It
should be stressed that this large freedom does not alter the physical
results, as long as all degrees of freedom are treated exactly. However,
in any approximate treatment, this may introduce ambiguities that should
be handled carefully, as we discuss below.

One salient feature of the JW transformation is that no constraint is
needed on the new variables (cf.\ for instance the Holstein-Primakoff or
Schwinger bosons), but non-locality is the main stumbling block in the
approach.

The success of the JW transformation in $1$ spatial dimension, in spite of
being non-local, resides on the fact that $XY$ nearest neighbors (NN)
interactions become local in fermion variables; this is not the case in
$2$ dimensions. Indeed, consider the $XY$ Hamiltonian on a given 2D
lattice
\beq
    H_{XY} = J \sum_{<p,q>} \left(S_p^x S_q^x + S_p^y S_q^y  \right) ,
\label{XY}
\eeq
where $J$ is the exchange constant and the sum runs
over all nearest neighbors $<p,q>$ on the lattice. In terms of fermion
variables the Hamiltonian reads
\beq
    H_{XY} = J \sum_{<p,q>} \left(\frac{1}{2}c^\dagger_p e^{i
\hat{\alpha}(p,q)} c_q + H.c. \right),
\label{XYJW}
\eeq
where
\beq
    \hat{\alpha}(p,q)= \sum'_r (\theta_{rq} - \theta_{rp} )c^\dagger_r c_r
\label{alpha}
\eeq
(the $'$ on the sum indicates that $\theta_{rr}$ terms are absent).
This phase is highly non local; in the 1D case,
this same expression becomes local due to the fact that the only two
actual values for the angles are $0$ and $\pi$. The non-locality in 2D is
usually overtaken by the introduction of an auxiliary gauge field
$A_{\mu}$, which on the one hand represents the phases in eq.(\ref{XYJW})
as the usual minimal coupling on the lattice, and on the other hand is
governed by a Chern-Simons action. The Gauss law associated to the first
order Chern-Simons action imposes a constraint which in anyon language attaches half a
quantum flux to each fermion, providing the statistical transmutation of
fermions into bosons. Then, a mean field treatment (known as {\it average
field approximation}) of the gauge field can be done, leading in general
to a quadratic NN interaction between fermions\cite{ARF}. However, the
Chern-Simons approach has serious difficulties when one deals with
arbitrary lattice topologies (for example the triangular lattice), and the
associated mathematical problems are not yet solved.

We do not introduce such an auxiliary gauge field, but keep working with fermion
variables. In order to perform numeric computations, one has to set a finite size lattice
and impose suitable boundary conditions. We use periodic boundary conditions, this
leading to a lattice on the torus. Moreover, the lattice size should be compatible with
possible periodic configurations; in the case of a square lattice, size must be even
in order to not interfere with the possible N\'eel order.

Now, it is not straightforward to define the JW transformation on the torus
\cite{Fradkin-Wen}, as the vector joining two different points is not unique . As one has
to take care of condition (\ref{key}), the vectors joining $p$ with $r$ and $r$ with $p$
must have arguments differing in $\pi$. We have to choose a unique segment joining each
pair of points $p,r$, and then draw both vectors along it. One can choose this segment by
a criteria of minimal distance. However, there exist pairs of points on the torus that
can be joined by two or more different segments with minimal distance and hence an {\it
ad-hoc} criterion must be added. Any such criterion unavoidably breaks translation
invariance, by preferring one segment over the rest. Naturally, we propose a criterion
trying to minimize the violation of translation symmetry as follows: we set a principal
finite size lattice and extend it on a plane by periodicity; for each point on the principal lattice
we consider also its periodic copies. Now, given a pair of sites, we look for the
shortest segment joining either the points or their copies; when such a segment is
unique, the procedure is translationally invariant.
For those pair of points where one can find more than one minimal distance segments, we
choose the one with both ends belonging to the principal lattice, thus breaking translation invariance.
Finally, the angles $\theta_{pr}$ and
$\theta_{rp}$ are computed as the arguments of the vectors joining $p$ and $r$ along the
chosen segment.
For convenience we also define that $\theta_{pp}=0$, in order to handle
the restriction on the sums in eqs.(\ref{jw}, \ref{alpha}).

As we mentioned in the Introduction, the JW transformation is exact but
the resulting Hamiltonian is
highly non-local and some kind of approximation is necessary to proceed.

We propose here a variational approach to deal with the non-local phases in eq.(\ref{XYJW})
and the quartic terms that can arise from $S^z$ interactions.
Working directly with fermion variables, we
replace the local fermionic occupation numbers $\hat n_p = c^\dagger_p
c_p$ by their expectation values in an arbitrarily chosen
variational state. This procedure leads to a multi-parameter
mean field approach, that will in turn be evaluated self-consistently.
This is the subject of the next section.

\section{Variational approach, applied to the $XXZ$ model}

To describe in full detail the method laid down above, we apply it to a generalized quantum spin 1/2 Heisenberg
antiferromagnet in a square 2D periodic lattice,
defined by the Hamiltonian
\beq
    \!\!\!\!
    H_{XXZ} = J \sum_{<p,q>} \left(S_p^x S_q^x + S_p^y S_q^y + \Delta
S_p^z S_q^z \right) -
    h  \sum_p S^z_p,
    \label{Heisenberg}
\eeq
where $\vec S_p=(S_p^x,S_p^y,S_p^z)$
represents the $S=1/2$ spin operator at site $p$, $J>0$ is the exchange
constant and $0 < \Delta < \infty$ the ``$XXZ$" anisotropy parameter. The
first sum in (\ref{Heisenberg}) runs over all nearest neighbors in the
given lattice, while the last term represents the interaction with a transverse external
magnetic field $h$. We work on a periodic rectangular lattice of size $K=N_x \times N_y$.

Using the JW transformation defined in eq.(\ref{jw}), the Hamiltonian can be
written in terms of spinless fermions as
\beq
    H_{XXZ} = J \sum_{<p,q>} \left[\frac{1}{2}
    (c^\dagger_p e^{i \hat\alpha(p,q)} c_q + H.c. ) +
    \Delta (c^\dagger_p c_p -\frac{1}{2}) (c^\dagger_q c_q -\frac{1}{2})
\right]-
    h \sum_p (c^\dagger_p c_p -\frac{1}{2} ) \ ,
    \label{hamjw}
\eeq
where the phase $\hat\alpha(p,q)$ is defined in (\ref{alpha}). Notice that the magnetic
field $h$ plays the r\^{o}le of a chemical potential for the JW fermions. In
particular,
we look for the ground state of the system (\ref{hamjw}) with fixed global magnetization
$M=0$ (corresponding to $h=0$).

We implement a self consistent mean field solution by starting with a given fermion
distribution profile $\{n_p\}$t (which can be random or guided by some ansatz) on the
lattice,
\beq
    \langle \hat n_p\rangle = n_p,
    \label{n_inicial}
\eeq
which has to satisfy a global constraint to
provide the given magnetization (here $\sum n_p =K/2$ corresponds to $M=0$).  We then replace the
operator $\hat \alpha(p)$ by its expectation value
\beq
    \langle \hat \alpha(p,q) \rangle =
    \sum_r (\theta_{rq}-\theta_{rp} ) \ n_r ,
    \label{MFphase}
\eeq
where the angles $\theta_{pq}$ are assigned following the criterion presented in the
previous section. To be precise, the principal lattice can be defined by indexing each site
by a position pair $(i,j)$, and setting the range $i=0 \cdots N_x-1$, $j=0 \cdots N_y-1$.
Periodic boundary conditions are then expressed by $ (i,j) \equiv (i + N_x, j)\equiv (i,
j + N_y) $.

Regarding the Ising term
\beq
    S_p^z S_q^z = c^\dagger_p c_p c^\dagger_q c_q -\frac{1}{2} c^\dagger_p c_p -
    \frac{1}{2}c^\dagger_q c_q -\frac{1}{4}
    \label{SzSz}
\eeq
in eq.(\ref{hamjw}), it is quartic in fermion operators, so we also treat it in mean
field. In order to approximate the first term in (\ref{SzSz}) with a quadratic expression
we propose the following
\beq
  c^\dagger_p c_p c^\dagger_q c_q \to \frac{1}{2}(c^\dagger_p c_p \langle
c^\dagger_q c_q \rangle +
    c^\dagger_q c_q \langle c^\dagger_p c_p \rangle),
    \label{SzMF}
\eeq
supported by best results in {\it a posteriori} evaluation of the GS energy (some other
possibilities are discussed in \cite{Derzhko}).

At this step, the Hamiltonian can be written as
\beq
    H_{XXZ}^{(MF)}(\{n_p\}) = \sum_{p,q} c^\dagger_p J_{pq}(\{n_p\}) c_q
    + C
    \label{totalMF}
\eeq
where
\beq
J_{pq} =
\left\{
    \begin{array}{lr}
        \frac{J}{2} \, e^{i\alpha(p,q)} &  {\rm if} <p,q> {\rm nearest \,
        neighbors}\\[2mm]
        \frac{J \Delta}{2} (\sum_{neighbors\ \  q} \langle n_q \rangle - 4 ) &  {\rm if \,} p=q
        \\[2mm]
        0 &  {\rm otherwise}
    \end{array}
\right.
\label{Js}
\eeq
and $C = KJ\Delta/2$.

The main idea of the present paper is to provide a systematic way to compute an
approximation to the true GS. We first find the GS for the quadratic
$H_{XXZ}^{(MF)}(\{n_p\})$ by solving the one particle (1P) spectrum and filling the
system with the lowest energy 1P states, up to the proper filling fixed by the total
magnetization $M$. Then we compute from this approximate GS a new set of local densities $n'_p =
\langle GS \vert c^\dagger_p c_p \vert GS \rangle $, which we use as a new input in
(\ref{totalMF}) and iterate this procedure looking for a fixed point configuration for
the density profile, {\it i.e.} a set of local densities $\{ n^*_p \}$ satisfying

\beq
n'_p(\{n^*_q\})= n^*_p \ .
\label{fpeq}
\eeq

The existence of a fixed point solution for this mapping
and its eventual dependence on a given initial configuration is not at all
obvious and has to be studied numerically.

In order to proceed with the method, $H_{XXZ}^{(MF)}(\{n_p\})$ can be
written in diagonal form
\beq
    H_{XXZ}^{(MF)}(\{n_p\})=\sum_{k=1}^K \epsilon(k) \ d^\dagger_k d_k +
{\rm constant}
    \label{XXdiag} \eeq
where $\epsilon_k$ are the 1P
eigenvalues of the quadratic part of $H_{XXZ}^{(MF)}$. Notice that $k$ is
just
an integer index over the spectrum, not to be confused with the lattice
momentum. Moreover, we order the eigenvalues ascendently.

The operators $d_k$ are related to $c_p$ by
\beq
    d_k=\sum_p Q^*_{kp} c_p
    \label{d}
\eeq
where $Q_{pk}$ is the matrix of
eigenvectors of $J_{pq}$. We compute both $\epsilon_k$ and $Q_{pk}$
numerically. Being $Q$ unitary, the set of $d_k$ operators satisfy fermion
commutation relations, $\{d_k,d^\dagger_{k'}\}=\delta_{kk'}$. Moreover,
the total fermion number operator satisfies
\beq
    N=\sum_{p=1}^K c^\dagger_p c_p = \sum_{k=1}^K d^\dagger_k d_k,
    \label{number}
\eeq
making it easy to control the filling in terms of the new fermions.

We now construct the approximation to the quantum GS as the half-filled
state that minimizes the energy, namely
\beq
    |GS \rangle = \prod_{k=1}^{K/2}  d^\dagger_k |0 \rangle
    \label{GS}
\eeq
Notice that this is a well defined quantum state of $K/2$ particles, except for casual
degeneracy of the 1P spectrum at the Fermi level . This is not the case for the $XXZ$
model on the square lattice (see details below).

From $|GS \rangle$ it is now easy to compute the approximate GS energy, as
\beq
    E_{GS} = \langle GS | H_{XXZ}^{(MF)} | GS \rangle = \sum_{k<K/2}
\epsilon_k +C.
    \label{EGS}
\eeq
Also the local occupation numbers can be computed in this approximate GS as
\beq
    n'_p = \langle GS | c^\dagger_p   c_p | GS \rangle
= \sum_{k<K/2} Q^*_{pk} Q_{pk}.
    \label{newns}
\eeq
With these occupation numbers we start again the procedure: compute
$J_{pq}$ in MF, diagonalize the new $H_{XXZ}^{(MF)}$, etc.

We have found after thorough numerical investigations that a fixed point solution for
eq.(\ref{fpeq}) always exists, but metastable solutions can also appear, depending on the
initial configuration one chooses. In any case, one can distinguish metastable solutions
from the best GS approximation simply by comparing their energies. Moreover, we describe
below how this drawback can be naturally solved by introducing a thermal bath to kick the
system out from the vicinity of metastable states.

Indeed, one can consider the effects of finite temperature by replacing
the proposed ground state (\ref{GS}) by a thermal state
$|\Psi_\beta\rangle$, compatible with the Fermi-Dirac 1P energy
distribution at a given temperature,

\beq
n(\epsilon)= \frac{1}{1+\exp(\beta (\epsilon - \bar\epsilon))},
\label{Fermi}
\eeq
where $\bar\epsilon$ is the 1P Fermi energy at half filling, and $\beta =
1/kT$ is the inverse temperature.  In detail, this thermal state
$|\Psi_\beta\rangle$ is constructed as

\beq
|\Psi_\beta\rangle= \prod_{k \in \sigma}  d^\dagger_k |0 \rangle,
\label{thermal}
\eeq
where $\sigma$ is a set of $K/2$ 1P states chosen with probability
$n(\epsilon(k))$ from some random simulation.

An exploration of the Hilbert space of the system by constructing a
thermal state from a starting fermion distribution, computing from it the
new local fermion distribution and again constructing a thermal state
should be considered as a thermalization at the given temperature. It
provides a source of thermal noise that has proven to help the system in
finding lower energy fixed points.

The thermalization can be done through several steps at a given
temperature, and then quenching to the pure quantum regime ($T=0$), or it
can be implemented by gradually lowering $T$ (annealing).

Besides, results at finite $T$ can also be achieved, by constructing an
statistical ensemble of microscopic states compatible with $T$.
Observables should then be computed as averages over the statistical
ensemble. We do not attempt to complete this program in the present
paper.

\section{Results}

We have tested the iterative approach described in the previous section
with the well known anisotropic $XXZ$ model on periodic 2D square lattices of size up
to $20 \times 20$ sites, at zero total magnetization. The sizes of the lattice that we explored
are by no means an upper limit, as our computations were made on a modest computer.
The anisotropy
parameter $\Delta$ has been explored in a range from $0.05$ to $1.5$,
including the isotropic $SU(2)$ case ($\Delta=1$, Heisenberg model).  As
starting configurations $\{n_p\}$ we have used random, uniform, and
different amplitude staggered distributions. We performed several
iterations and analyzed the evolution of the local fermion profile and the
approximate GS energy. We report the results in terms of spin variables,
noting that the local fermion occupation represents the local
magnetization as $m_z(p)= (n_p-\frac{1}{2})$.

\begin{figure}[htbp]
\begin{center}
\epsfig{file=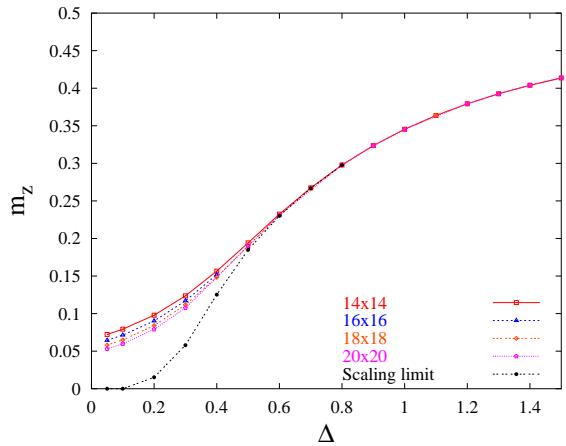,width=6cm,angle=270}
\end{center}
\vskip .5 truecm
\caption{N\'eel order parameter at fixed points, in function of the anisotropy $\Delta$.
Several lattice sizes and the scaling limit are shown.}
\label{f1}
\end{figure}

\begin{figure}[htbp!]
\begin{center}
\epsfig{file=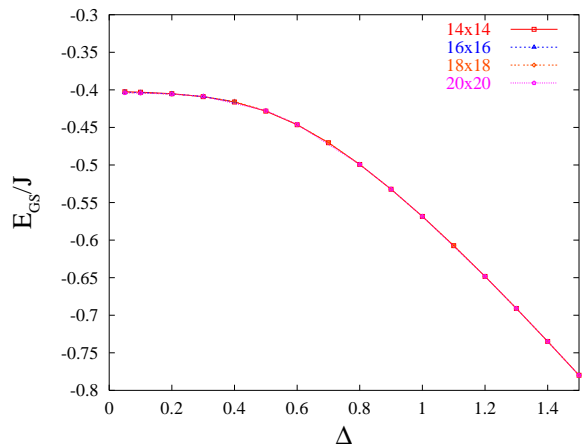,width=6cm,angle=270}
\end{center}
\vskip .5 truecm
\caption{GS energy, in function of the anisotropy $\Delta$. Data
corresponds to configurations plotted in figure (\ref{f1}).}
\label{f2}
\end{figure}

Working at $T=0$, we have found that in general, from different starting configurations,
the system rapidly finds a N\'eel order as stable ground state approximation, after $15
\sim 20$ iterations. The N\'eel order parameter, usually defined as the staggered or
sublattice magnetization $m_z$, depends on the anisotropy parameter $\Delta$. Fluctuations
around this staggered magnetization are typically of order $10^{-8}$. In figure
(\ref{f1}) we plot the N\'eel order parameter $m_z$ of the fixed point solution for
different values of $\Delta$, for several lattice sizes.
Finite size effects are
noticeable for lower values of $\Delta$, so we also show the results of a finite size
scaling $m_z(\infty)$ of our data, fitted with a power law $m_z(K)=m_z(\infty)+
c/K^\alpha$. The corresponding GS energies per site are shown in figure (\ref{f2}) where
one observes that scaling with the system size is clearly less important.
We have observed that the 1P spectrum of the mean field Hamiltonian (\ref{totalMF})
presents a gap $2 m_z J \Delta$ for N\'eel ordered configurations, at the half filling
Fermi level. This is in agreement with (\cite{ARF}) and makes the construction of
$|GS\rangle$  in eq.(\ref{GS}) unambiguous.

\begin{figure}[htbp!]
    \begin{center}
        \epsfig{file=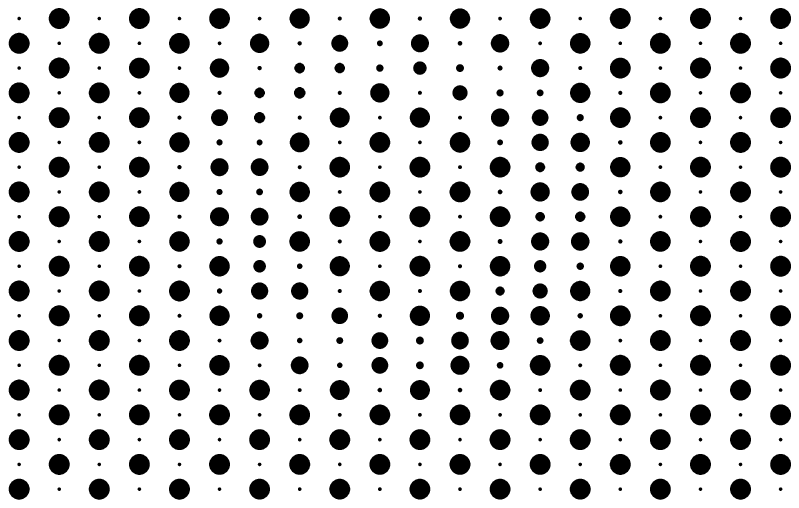,width=6cm,height=5cm,angle=0} \hspace{1.5cm}
        \epsfig{file=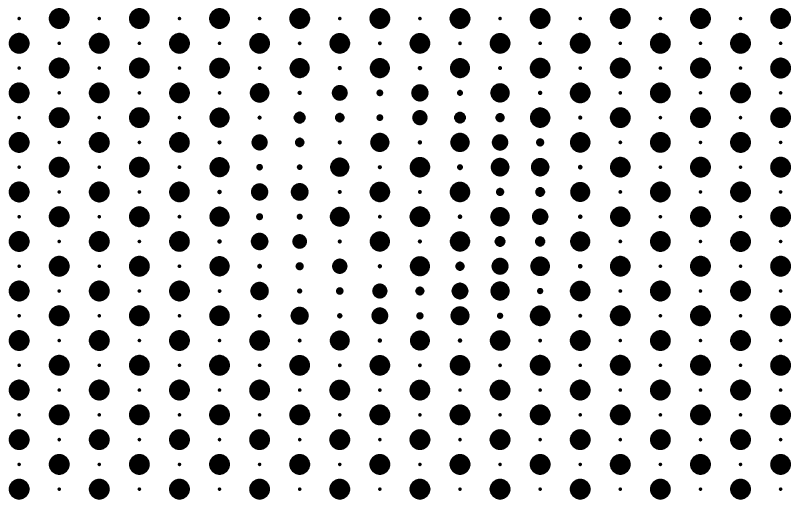,width=6cm,height=5cm,angle=0}
    \end{center}
    \vskip .5 truecm
    \caption{Occupation patterns for a metastable configuration, where antiferromagnetic domains
    appear. The size of the points is proportional to the
    local fermion occupation number. This configurations occurred on a $20\times 20$ lattice,
    with $\Delta=1.1$, after 10 steps of thermalization at $T=0.2~J$, and 10 (left panel) or 20 (right panel)
    more steps of GS search at $T=0$. The smaller domain is seen to decrease in size under the simulated evolution.}
\label{f3}
\end{figure}

In the case of random initial distributions, metastable configurations can show up; a
detailed inspection of the local magnetization in these cases reveals the formation of
antiferromagnetic domains, that is the presence of the two possible N\'eel configurations
in different regions. In figure (\ref{f3}) we show an example of such domains, at two
different stages of a sample evolution. It is natural to expect that larger lattices
favor the formation of these domains, as it indeed is observed. These configurations have
higher energy than the uniform N\'eel state and correspond then to metastable
configurations; correspondingly, they are not presented in figures (\ref{f1}, \ref{f2}).

\begin{figure}[htbp!]
    \begin{center}
        \epsfig{file=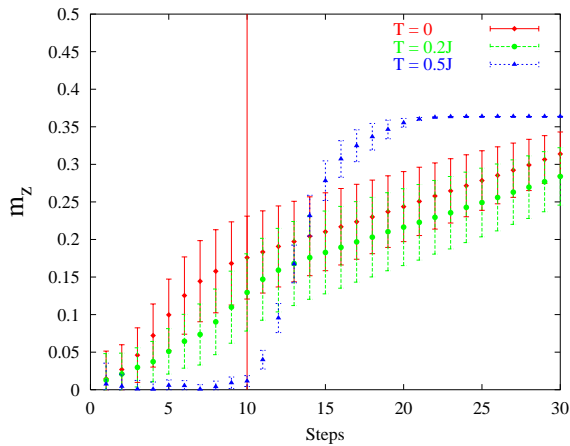,width=6cm,angle=270}
    \end{center}
    \vskip .5 truecm
    \caption{Example of the evolution of the N\'eel order parameter from
    a sample initial random configuration. Data corresponds to the system depicted in fig.(\ref{f3}),
    with a vertical line separating the thermal evolution and the $T=0$ evolution.
    Error bars indicate the standard deviation of local magnetization from N\'eel order (reduced by a factor of 5 for clarity).
    Insufficient thermalization can lead to metastable configurations or to
    very slow convergence, while higher temperature dramatically improves convergence towards an ordered configuration.}
\label{f4}
\end{figure}

When a thermal bath is simulated on random initial configurations, we have
observed that metastable configurations are less likely to appear. After
thermalization we let the system to cool down by either quenching or
annealing as described in Section III, and complete the iterations at
$T=0$. In fact, a few steps ($\sim 10$) of thermalization with
sufficiently high $T$ completely avoid domain formation and lead to a
unique fixed point mean field configuration; the required temperature is
higher for larger lattices, being of the order of $J$ for the lattice of
$20 \times 20$.
We have checked that under general circumstances, quenching provides the fastest
convergence method to the minimum energy state.
An example of the evolution of the N\'eel order parameter from an
initial random configuration, under thermalization with different temperatures, is shown in
figure (\ref{f4}).

The results of the present MF computation show all the features expected for 
the Heisenberg
antiferromagnet on the square lattice. They are of course not comparable to accurate
numerical techniques \cite{Cuccoli}, but are in qualitative agreement with results from
previous studies. In particular, in the scaling limit we obtain no N\'eel order for small
anisotropy $\Delta$, where the system presumably has $XY$ order. We can estimate a
critical value $\Delta^* \approx 0.2$, above which N\'eel order develops.  For the
isotropic Heisenberg point $\Delta=1$ we obtain a sublattice magnetization $m_z=0.3453$,
with ground state energy per site $E_{GS}/K= -0.5683 J$, to be compared for instance with
corresponding Quantum Monte Carlo values of $0.307$ and $-0.6694 J$ \cite{QMC}.

\section{Conclusions}

We have presented a self consistent MF procedure for exploring the quantum
ground state of any $S=1/2$ spin system on a 2D lattice. When tested on
the $XXZ$ model on a square lattice, the method provides the correct
qualitative description of the system, with no {\em a priori} ansatz for
any kind of order. We computed the values for the sublattice magnetization
and GS energy for a wide range of values of the anisotropy parameter,
which compare qualitatively well with the available numerical data, at
least for $\Delta = 1$ where most accurate data is available. Moreover, we have found
that the sublattice magnetization as a function of the $XXZ$ anisotropy
shows the correct qualitative behaviour, expected from a spin wave
analysis \cite{ARF}.

The present approach has a more general scope than previous MF
computations, in the sense that it can be applied to
{\it any} lattice topology, irrespectively of the appearance of
frustrating units, a fact that prevents the applicability of one of
the most powerful numerical techniques such as Quantum Monte Carlo. A
magnetic field can be trivially added as a chemical potential for the JW
fermions and hence magnetization curves could be obtained. Since the
method is not based on any periodicity of couplings, it can be well suited
to study disordered quantum spin systems, at the only price of increasing
the CPU time. Last but not least, the approach is naturally well suited
for the study of the thermodynamics of these systems, since temperature
can be added in a simple way.

Among other situations, it would be interesting to apply this technique to the Heisenberg
quantum AF on the triangular lattice, where there is disagreement between Chern-Simons MF
predictions \cite{Misguich} and numerical data about a magnetization plateaux at zero
magnetization. Another case of interest the kagom\'e lattice, where a quantum spin liquid
is believed to be realized \cite{Lluillier} (see also \cite{ojo}). This issues will be
investigated elsewhere.

\vspace{1cm}

{\em Acknowledgements}: We are especially grateful to M.\ Grynberg and A.\
Honecker for useful discussions and computational help.  We also thank C.\
Balseiro, W.\ Brenig, J.\ Drut and E.\ Fradkin for useful comments.
We acknowledge CONICET and Fundaci\'on Antorchas
(grants No.\ 14116-11 and 14022-79) for financial support, and the Ecole 
Normale Sup\'erieure de Lyon, where part of this work was done.


\begin{thebibliography}{99}

\bibitem{Anderson}
P.W.\ Anderson, Science {\bf 235}, 1196 (1987).

\bibitem{SrCuBo} K.\ Onizuka {\it et al.}, J.\ Phys.\ Soc.\ Jpn.\
{\bf 69}, 1016 (2000).

\bibitem{JW}
P.\ Jordan, E.P.\ Wigner, Z.\ Phys.\ {\bf 47}, 631 (1928).

\bibitem{Fradkin}
E.\ Fradkin, Phys.\ Rev.\ Lett.\ {\bf 63}, 322 (1989).

\bibitem{ARF}
A.\ Lopez, A.G.\ Rojo, E.\ Fradkin, Phys.\ Rev.\ B {\bf 49}, 15139
(1994).

\bibitem{Wang} Y.R.\ Wang, Phys.\ Rev.\ B {\bf 43}, 3786
(1991); {\bf 45}, 12604 (1992); {\bf 45}, 12608 (1992);
K.\ Yang, L.K.\ Warman, S.M.\ Girvin, Phys.\ Rev.\ Lett.\ {\bf 70}, 2641
(1993).

\bibitem{Derzhko} O.\ Derzhko, J.\ Phys.\ Studies (L'viv) {\bf 5}, 49 (2001).

\bibitem{Misguich}  G.\ Misguich, Th.\ Jollicoeur, S.M.\ Girvin,
Phys.\ Rev.\ Lett.\ {\bf 87}, 097203 (2001).

\bibitem{Chang} M-C.\ Chang, M-F.\ Yang, Phys.\ Rev.\ {\bf B66}, 184416 (2002).

\bibitem{Derzhko2003} O.\ Derzhko, T.\ Verkholyak, R.\ Schmidt, J.\ Richter,
Physica {\bf A320}, 407 (2003).

\bibitem{Semenoff} J.\ Ambjorn, G.\ Semenoff, Phys.\ Lett.\ {\bf B226},
107 (1989).

\bibitem{Fradkin-Wen} Related problems in the gauge field approach
were discussed in X.G.\ Wen, E.\ Dagotto, E.\ Fradkin, Phys.\ Rev.\ B
{\bf 42}, 6110 (1990).

\bibitem{Cuccoli} A.\ Cuccoli, T.\ Roscilde, V.\ Tognetti, R.\ Vaia, P.\ Verruchi, J.\ Appl.\ Phys.\ {bf 93}, 7640 (2003).

\bibitem{QMC}  M.\ Calandra Buonaura, S.\ Sorella, Phys.\ Rev.\ B {\bf 57},
11446 (1998).

\bibitem{Lluillier} Ch.\ Waldtmann, H.-U.\ Everts, B.\ Bernu, P.\ Sindzingre,
C.\ Lhuillier, P.\ Lecheminant, L.\ Pierre, Eur.\ Phys.\ J.\ B {\bf 2},
501 (1998)

\bibitem{ojo} P.\ Nikolic, T.\ Senthil, preprint cond-mat/0305189.

\end{thebibliography}
\end{document}